\documentclass{article}
\PassOptionsToPackage{numbers, compress}{natbib}

\usepackage[preprint]{neurips_2025}

\usepackage[utf8]{inputenc} 
\usepackage[T1]{fontenc}    
\usepackage{hyperref}       
\usepackage{url}            
\usepackage{booktabs}       
\usepackage{amsfonts}       
\usepackage{nicefrac}       
\usepackage{microtype}      
\usepackage{xcolor}         
\usepackage{graphicx} 

\title{Modernizing Facebook Scoped Search: Keyword and Embedding Hybrid Retrieval with LLM Evaluation}

\author{%
  Yongye Su\thanks{Work done at Meta.} \\
  Meta\\
  Menlo Park, CA\\
  \texttt{yongye@meta.com} \\
  \And
  Zeya Zhang\\
  Meta\\
  Bellevue, WA\\
  \texttt{zeyaz@meta.com} \\
  \And
  Jane Kou\\
  Meta\\
  Bellevue, WA\\
  \texttt{janekou@meta.com} \\
  \AND
  Cheng Ju\\
  Meta\\
  Menlo Park, CA\\
  \texttt{jucheng@meta.com} \\
  \And
  Shubhojeet Sarkar\\
  Meta\\
  Menlo Park, CA\\
  \texttt{shubhojeet@meta.com} \\
  \And
  Yamin Wang\\
  Meta\\
  Bellevue, WA\\
  \texttt{yaminwang@meta.com} \\
  \And
  Ji Liu\\
  Meta\\
  Bellevue, WA\\
  \texttt{madisonliu@meta.com} \\
  \And
  Shengbo Guo\\
  Meta\\
  Menlo Park, CA\\
  \texttt{shengbo@meta.com} \\
}

\begin{document}

\maketitle

\begin{abstract}
  Beyond general web-scale search, social network search uniquely enables users to retrieve information and discover potential connections within their social context. We introduce a framework of modernized Facebook Group Scoped Search by blending traditional keyword-based retrieval with embedding-based retrieval (EBR) to improve the search relevance and diversity of search results. Our system integrates semantic retrieval into the existing keyword search pipeline, enabling users to discover more contextually relevant group posts. To rigorously assess the impact of this blended approach, we introduce a novel evaluation framework that leverages large language models (LLMs) to perform offline relevance assessments, providing scalable and consistent quality benchmarks. Our results demonstrate that the blended retrieval system significantly enhances user engagement and search quality, as validated by both online metrics and LLM-based evaluation. This work offers practical insights for deploying and evaluating advanced retrieval systems in large-scale, real-world social platforms.

\end{abstract}

\section{Introduction}

Facebook Search operates as one of the world's largest social search engines, serving billions of queries daily across diverse content types including people, pages, groups, events, and posts~\cite{meta_fb_search, 10.1145/3447548.3467127, curtiss2013unicorn}. Aside from general platform search, Facebook Group Scoped Search enables users to efficiently find relevant information and re-discover posts within specific social or geological communities. However, existing legacy keyword-based retrieval systems are only able to capture exact keyword-matched cases of candidate posts, but are unable to capture semantic relationships between queries and content, particularly for natural language inquiries that may not contain exact keyword matches. For example, a query like "small individual cakes with frosting" would miss posts about "cupcakes" group post. With current keyword-only search, users would only see posts that literally contain their descriptive phrases, potentially missing highly relevant content that uses the proper terminology or common names. On the other hand, group scoped search is a filtered query that has much less base content compared with global search, keyword-only retrieval leads to less search results, potentially causing less engagement.

The challenges of introducing semantic search are considerable. First, adding embedding-based retrieval (EBR)~\cite{10.1145/3394486.3403305, 10.1145/3447548.3467127, karpukhin2020dense, xiong2020approximate, khattab2020colbert} introduces substantial computational overhead, general semantic retrieval model often requires GPU inference for turning natural language query into embedding representation and increases latency by a few milliseconds compared to traditional keyword search~\cite{nogueira2019passage}. Second, unlike keyword-based retrieval which has clear explainability (you can see exactly which keywords matched), EBR retrieves content based on similarity scores in high-dimensional embedding space. For example, searching for "Italian coffee drink" might return posts about "cappuccino" with a similarity score of 0.89, but only people with contextual knowledge can understand why there's a match, making it challenging to evaluate EBR results rapidly at scale~\cite{yang2018anserini}. Third, the integration complexity of blending dense semantic retrieval with existing sparse keyword systems also requires modernized ranking and score fusion mechanisms~\cite{kuzi2020leveraging, formal2021splade}.

To address these systems challenges, we present a modernized approach that extends Facebook's proven EBR architecture~\cite{10.1145/3394486.3403305} to group scoped search, leveraging both lexical precision and semantic understanding while maintaining production-level performance requirements~\cite{malkov2018efficient}. The goal of the system is to render more semantically-relevant and diverse group posts within the group. To rigorously evaluate improvements, we introduce a scalable, LLM-based build verification test (BVT) framework for automated group scoped search relevance assessment, enabling rapid iteration and robust quality measurement.


\begin{figure}[h]\label{fig:architecture}
    \centering
    \includegraphics[width=1\linewidth]{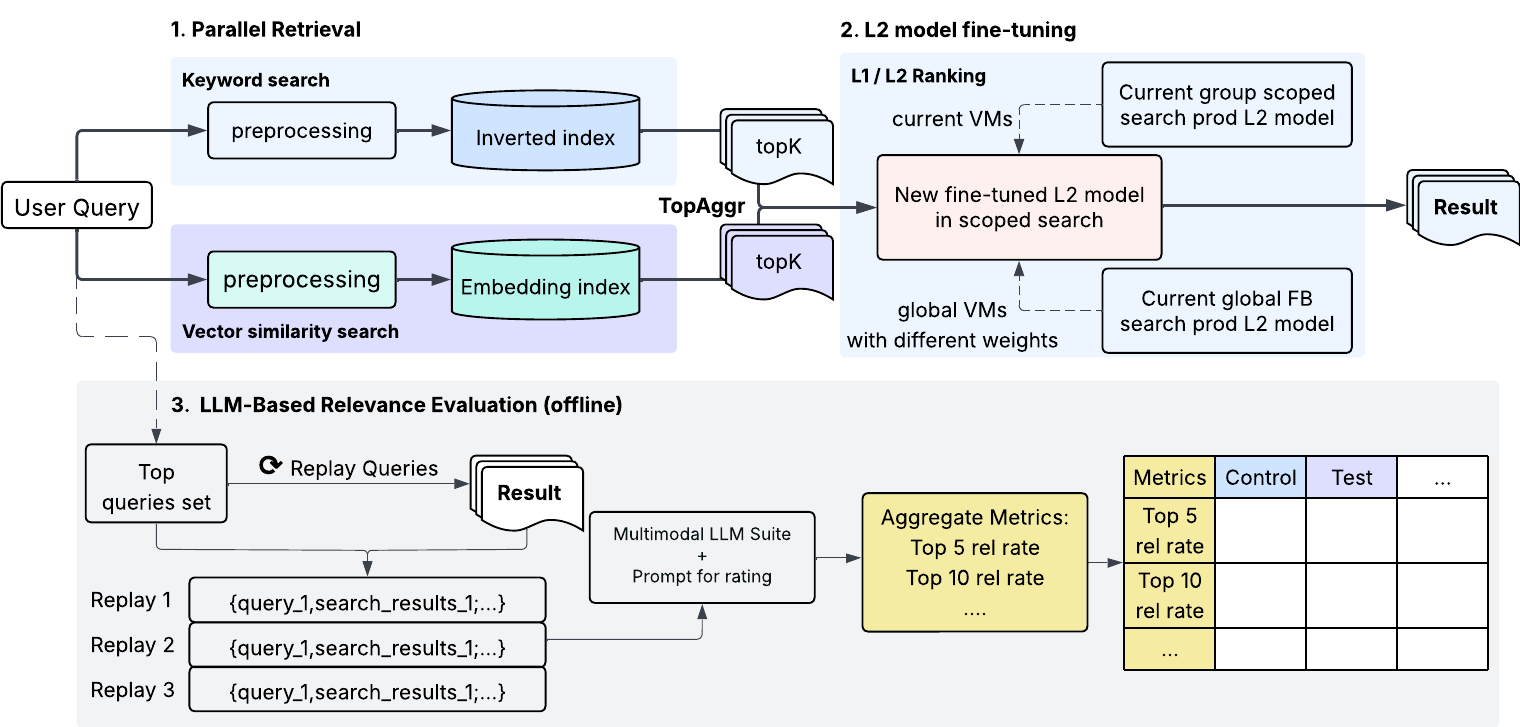} 
\caption{System architecture for blended retrieval in Facebook Group Scoped Search with LLM-based search relevance evaluation. Part 1 (Parallel Retrieval) corresponds to Retrieval Stage Design and Part 2 (L2 model fine- tuning) represents Ranking Stage Design.}
\end{figure}

\section{System Design}

\subsection{Hybrid Retrieval Stage Design}

To modernize Facebook Group Scoped Search, we introduce a blended retrieval architecture as shown in the retrieval part of Figure~\ref{fig:architecture} that unifies traditional keyword-based search with embedding-based retrieval (EBR), also known as Search Semantic Retriever (SSR). This design extends the complementary strengths of both lexical precision and semantic understanding, enabling the system to render more semantically-relevant and diverse group posts content, especially for ambiguous or natural language sentence queries.

When a user submits a query within a group, before parallel retrieval, the system first applies comprehensive query preprocessing pipelines respectively for each retrieval method. This step includes tokenization, normalization, and, where appropriate, query rewriting or decomposition. Such preprocessing is essential not only for the legacy keyword-based retrieval system, which relies on clean and well-structured queries for effective inverted index matching, but also for the SSR embedding-based system. SSR is a 12-layer 200M parameter model, it is to generate meaningful embeddings that capture the user’s intent. Following preprocessing, the query is dispatched in parallel to both retrieval pathways. The keyword-based system uses Facebook’s Unicorn inverted index to fetch posts containing exact or closely matched terms, ensuring high precision for well-specified queries.  Simultaneously, the SSR model encodes the query into a dense vector representation. This embedding is then used to perform an approximate nearest neighbor (ANN) search over a precomputed Faiss vector index~\cite{douze2024faiss,johnson2019billion} of group post embeddings, which are generated offline by the document tower of the SSR model. 

From an infrastructure perspective, the EBR subrequest integrates into Facebook's existing search backend through targeted modifications to the Unicorn query processing pipeline~\cite{curtiss2013unicorn}, allowing for rapid experimentation and scalable deployment. Efficient preprocessing and optimized ANN algorithms are employed to ensure that embedding indices remain within infrastructure budgets and that query latency is minimized. The underlying distributed architecture and strategic caching further support high query throughput and low response times, making the system robust and scalable for Facebook’s large user base.

\subsection{Ranking Stage Design}

As shown in Figure~\ref{fig:architecture}, with new additional EBR results returned to the ranking stage, we also further fine-tune the ranking stage models to fit this upper stream modification. The candidates retrieved from both the keyword and embedding-based systems are then merged in such a candidate ranking stage, where each candidate is retrieved with lexical features (TF-IDF, BM25 scores) and/or semantic features (cosine similarity scores). The combined candidate set is subsequently passed through Facebook's multi-stage ranking pipeline. The current L2 model employs a single architecture with direct feature input leading to a single output score. However, our anticipated supermodel design introduces a combined architecture with sub-models and Multi-Task Multi-Label (MTML) supermodel components~\cite{meta_transparency_ranking}, enabling the system to jointly optimize for multiple engagement objectives while maintaining the plug-and-play modularity.

This plug-and-play L2 supermodel architecture enables easier maintenance and faster iteration cycles while targeting state-of-the-art performance for scoped search ranking. The ranking optimization focuses on three key user engagement signals with weighted importance: click, share, and comment, allowing the system to balance immediate relevance with long-term user engagement.

\subsection{LLM-Based Evaluation Design}

To ensure continuous improvement and maintain high relevance standards in Group Scoped Search, we implement an innovative LLM-based evaluation framework within its Build Verification Test (BVT) process as Part 3 of Figure~\ref{fig:architecture}. This LLM-as-a-Judge framework applies large language models, specifically, Llama 3 with multi-modal capability, as automated judges to assess the quality of search results at scale.

The evaluation process focuses on scoped query sets, by using these query sets in each corresponding groups, we get multiple query and simulated search results pairs from search engine result pages (SERPs) backend engine via multiple rounds of replaying. We also applies carefully designed prompt templates guide the LLMs to deliver reliable relevance scores based on multiple rounds of judgement, capturing degrees of relevance rather than single judgment to enhance stability and interpretability. This approach allows for a more refined understanding of search quality. Furthermore, the framework provides potential directions for ranking optimization, since evaluates results across multiple result set sizes, such as the top-k results (k=5, 10...), providing detailed insights into the search system’s performance at different depths. This LLM-based evaluation framework offers several key benefits. It enables rapid, large-scale evaluation of search system changes without the bottleneck of human labeling, significantly accelerating development cycles. The use of LLMs reduces variability and bias inherent in human judgments, providing stable and reliable quality signals. This consistency supports faster iteration by quickly surfacing regressions or improvements in search relevance. Additionally, the framework reduces evaluation expenses compared to manual annotation efforts.

For implementation, the BVT framework integrates seamlessly with Meta’s search infrastructure. It replays real search queries within group contexts and applies LLM-based relevance scoring to the returned results. This integration supports continuous monitoring and validation of search quality, ensuring that new features or model updates meet expected standards before deployment. Overall, this LLM-based evaluation framework represents a significant advancement in automated search quality assessment, enabling Meta to maintain high standards while supporting rapid innovation.

\section{Evaluation}

\subsection{Prompt Design and Evaluation Metrics}
We employ a structured prompt template approach to guide LLM-based relevance assessment, focusing on capturing nuanced relevance relationships in group contexts. Our evaluation framework tracks several key metrics to assess system performance comprehensively. The \textbf{relevant rate} measures the proportion of results judged as contextually relevant, providing a high-precision assessment of retrieval quality. The \textbf{somewhat relevant rate} measures the proportion of results judged as either relevant or somewhat relevant, providing a more inclusive assessment of retrieval quality. We define "somewhat relevant" through explicit prompt guidance: 
\textit{"If you are a rater evaluating whether a query is somewhat relevant to a post, consider cases where both share a common domain or theme. For example, if both the query and post are related to sports, then they are somewhat relevant, even if the specific sports differ.  Search Query: \{query\}, User: \{user\}"}
This relaxed relevance criterion captures semantically related content that traditional keyword matching might miss. The \textbf{error rate} quantifies instances where the LLM-as-a-Judge fails to provide valid assessments due to prompt misunderstanding, output formatting issues, or model limitations. The \textbf{skip rate} tracks cases where the evaluation framework cannot process certain query-result pairs, due to malformed inputs or technical constraints in the evaluation pipeline.

\subsection{Results and Discussion}
As shown in Table~\ref{tab:response_metric_comparison_updated}, the evaluation results demonstrate consistent improvements across all new components of proposed system, with the hybrid approach showing complementary benefits from both embedding-based retrieval and ranking model enhancements.

\begin{table}[!ht]
\centering
\caption{LLM Search Result Judgement Results}
\label{tab:response_metric_comparison_updated}
\begin{tabular}{lcccc}
\toprule
Response Metric & Baseline & New L2 model & EBR & New L2 + EBR \\
\midrule
top5 rel rate & 84.7 & 85.1 & 85.5 & 85.2 \\
top5 somewhat rel rate & 94.1 & 94.5 & 94.6 & 94.7 \\
top5 err rate & 10.8 & 10.8 & 10.8 & 10.4 \\
top5 skip rate & 2.1 & 1.9 & 1.9 & 2.0 \\

top10 rel rate & 0.9 & 0.902 & 0.901 & 0.904 \\
top10 somewhat rel rate & 0.966 & 0.967 & 0.969 & 0.97 \\
top10 err rate & 0.152 & 0.144 & 0.152 & 0.146 \\
top10 skip rate & 0.011 & 0.01 & 0.011 & 0.011 \\

\bottomrule
\end{tabular}
\end{table}

\section{Conclusion and Future Work}
In this work, we presented a modernized Facebook Group Scoped Search by integrating legacy keyword retrieval with embedding-based retrieval. To better accommodate such mixed retrieval results, we updated L2 ranking model with MTML architecture. We also demonstrated the effectiveness of this approach through large-scale offline LLM-as-a-judge testing, showing successful in search relevance and quality. Looking ahead, several promising directions can further enhance the system, such as applying LLMs in the ranking stage to further increase the result relevance, using LLM-driven adaptive retrieval strategies, and so on.

\section{Acknowledgement}
Authors hereby sincerely appreciate the team for the project design and their insightful feedback.

\bibliographystyle{plain}
\bibliography{neurips_2025}

\appendix


\end{document}